# Opioids for pain treatment of Cancer:

# a Knowledge Maturity Mapping

Cesar Aguado[1], J. Arturo Silva-Ordaz[2] and Victor M. Castaño[1]

1.-Universidad Nacional Autónoma de México

Ciudad Universitaria

Cd. de México, 04510 MEXICO

2.- Quantum Works, SC

Querétaro, MEXICO

**Abstract**

The conceptual structure of opioids, based on the bibliometric analysis of 4,935 articles of the Web of Science using the following indices: Science Citation Index Expanded (SCI-EXPANDED), Social Sciences Citation Index (SSCI), Conference Proceedings Citation Index - Science (CPCI-S), Conference Proceedings Citation Index- Social Science & Humanities (CPCI-SSH), Book Citation Index– Science (BKCI-S), Book Citation Index - Social Sciences & Humanities (BKCI-SSH) y Emerging Sources Citation Index (ESCI), was constructed. We analyzed the available articles with the words "Opioids" and "Cancer". We show the evolution and the state of the art in countries where these treatments are implemented. The results were processed identifying the most cited articles to extract the main connections and frequencies of key words, authors, journals, countries, institutions, and their tendencies and their connection and degree of collaboration. The temporal tendencies, the word cloud, the keyword network, the evolution of words, author's production and the scientific production by country are analyzed in terms of the increasing frequency in which opioids are employed to treat both cancerous and non-cancerous pain.

*Keywords: opioids; cancer; pain; bibliometrics.*

**Introduction**

Opioids constitute the most effective treatment against pain caused by cancer, that is why a careful evaluation of pain is needed [1]. Furthermore, opioids find increased use in the treatment of non-cancerous pain [2]. Table 1 show a classification of cancer pain:

Table 1

Classification of pain

| | |
|---|---|
| Cancer related | 1. Pain related to the illness: bone invasion, soft tissue, muscles, nervous structures, lymphedema, intestinal obstruction, and intracraneal pressure increase. |
| | 2. Pain related to the treatment: postsurgery, postchemotherapy, postradiation, and phantom pain. |
| | 3. General weakness and cancer progression: constipation and bed pain. |
| | 4. Other: migraine, back pain, arthritis |
| Nerve related | 1. Nociceptive pain (somatic or visceral pain): bone metastasis, bone fracture due to vertebral compression. |
| | 2. Neuropathic pain: pain induced by chemotherapy, post-herpes neuralgia, and phantom pain. |

In recent years, the prevalence of chronic pain has provoked an opioid epidemic with adverse consequences, such as the liberalization of the regulation that deal with the prescription of opioids as treatment of pain not related to cancer, and has led to a dramatic increase in opioid use [3].

The successful treatment of pain with opioids requires an adequate classification of analgesics without excessive adverse side effects. The management of adverse side effects is still a clinical challenge [4]. Table 2 shows a classification of opioid analgesics.

Table 2

Classification of analgesics

| Agonist | Partially agonist | Antagonist |
|---|---|---|
| Codeine | | |
| Tramadol | | |
| Hydrocodone | Pentazocine | |
| Morphine | Nalbuphine | Naloxone |
| Oxycodone | Buprenophine | |
| Hydromorphone | Butorphanol | |
| Fentanyl | | |

There is a controversy around the long-term use of opioids in non-cancerous pain patients[5]. There is also the concern of the problems related with addiction or abuse in patients under opioid therapy.

There is a contrast with the use of opioids in the treatment of acute and chronic pain related to cancer by patients who achieved partial relief of pain in comparison with chronic therapy with opioids for non-cancerous pain, which is still controversial [6]. The observation is that opioids diminish and improve the conditions for patients with cancer.

When pain is not treated and effectively relieved, there is a harmful effect in every aspect of the quality of life, and it has been shown that a correct analgesic therapy improves the quality of life, relieving the pain of cancer patients [7].

**Methodology**

The methodology used consists on identifying, from a database of 4,935 articles downloaded from the Web of Science via the bibliometric techniques used to relate the scientific work that has had the most impact in the research of opioids and cancer analyzing the intellectual structure in the research fields of the articles [8,9,10].

The bibliometric technique analyses the fields of the research articles including citations, impact indices, most relevant sources, concurrence of words and content by means of quantitative methods to explore the patterns, tendencies, relationships and networks.

We achieved this with the use of bibliometric tools, which also helped us to identify key research themes, and the interrelationship between opioids and cancer. We show a solid conceptual structure in this research field through indicators, graphs and tables.

The terms searched in the bibliometric analysis were "opioids" and "cancer", to show the intellectual structure through these scientific publications.

The Table 3 summarizes the main results related with the search of the terms "opioids" and "cancer".

Table 3

Sources for bibliometric analysis

| Description | Results |
| --- | --- |
| Articles | 4,935 |
| Sources (books, journals, etc.) | 1,136 |
| Period | 1985 - 2020 |
| Average number of citations per article | 25.11 |
| Authors per document | 3.12 |
| Coauthors per document | 4.93 |
| Collaboration index | 3.46 |

The search began with the keywords "opiods" and "cancer", and we obtained 4,935 published articles out of 75'184,964 existing articles in Web of Science.

**Results and discussion**

In Figure 1 we find that the tendency of these two subjects (opioids and cancer) has increased very rapidly between 1985 and 2019. In 1994 we find only 34 articles, 72 articles in 1997 and 108 articles in the year 2000. In 2019 there are 449 articles related to these subjects.

Figure 1

Annual Scientific Production

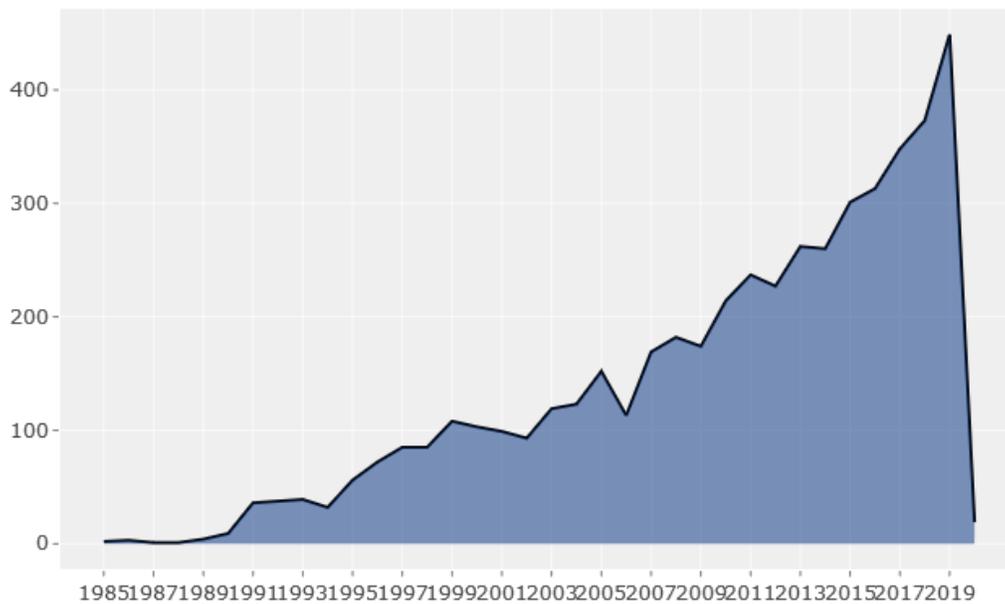

We found other important information in the most relevant sources, as shown in Table 4.

Table 4

Sources with greatest number of articles

| Source | Number of Published Articles |
|---|---|
| Journal of Pain and Symptom Management | 408 |
| Supportive Care in Cancer | 167 |
| Pain | 130 |
| Pain Medicine | 122 |
| Palliative Medicine | 103 |

The Table 5 shows the H, G and M indexes impact indexes of the journals with the greatest number of published articles.

Table 5

Journal impact

| Source | H Index | G Index | M Index | Total citations |
|---|---|---|---|---|
| Journal of Pain and Symptom Management | 68 | 99 | 2.26 | 15,637 |
| Supportive Care in Cancer | 30 | 47 | 1.07 | 3,194 |
| Pain | 55 | 101 | 1.71 | 10,391 |
| Pain Medicine | 31 | 48 | 1.47 | 2,900 |
| Palliative Medicine | 31 | 49 | 1.19 | 2,996 |

In Table 6 we can find the frequency of key words related with the search of "Opioids" and "Cancer".

Table 6

Keyword frequency

| Keyword | Frequency |
| --- | --- |
| Opioid | 1,161 |
| Cancer pain | 662 |
| Pain | 602 |
| Palliative care | 413 |
| Cancer | 400 |
| Chronic pain | 346 |
| Opioid | 334 |
| Morphine | 308 |
| Pain management | 195 |
| Analgesics | 188 |

Figure 2 shows that, in the 4,935 articles of our study, there is a relationship and a steady growth since 1990 in the articles with the keywords "opioids" with 1149 occurrences, "cancer pain" with 657 occurrences and "pain" with 597 occurrences.

Figure 2

Growth of keywords

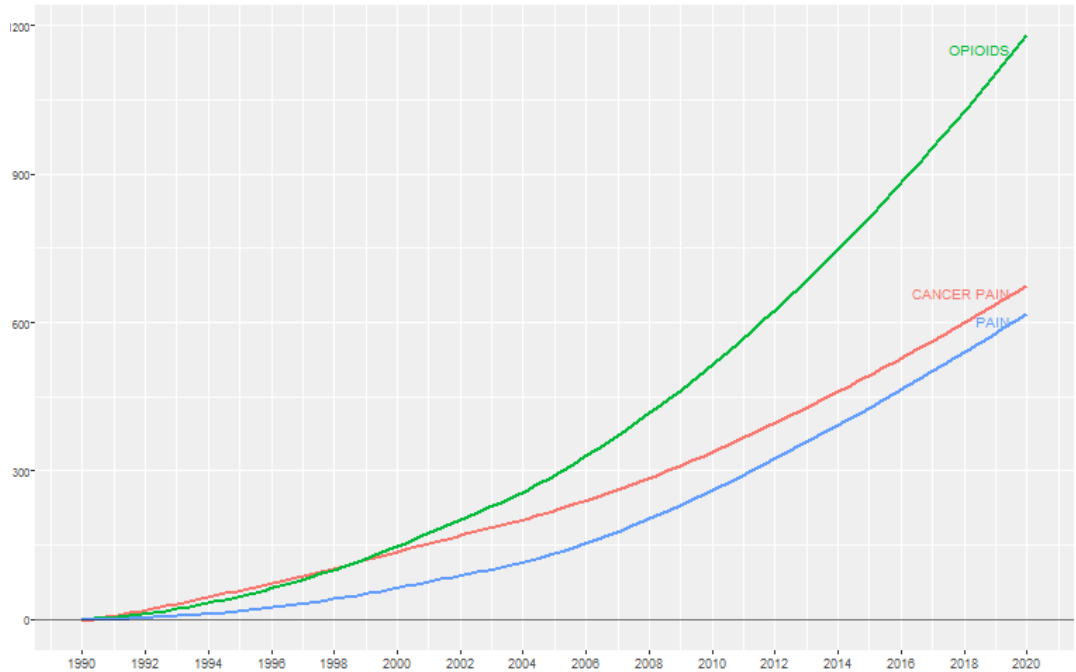

The bibliometric analysis shown in Figure 3 shows the relationship between keywords with the most prominent journals. This results in Table 6, that shows the clusters and keywords.

Figure 4 shows the relationship of the variables found in the articles. Two main ones are shown, on one side the connection of "opioids", "chronic pain", "pain management", and on the other, "cancer", "pain", "opioid", "morphine", "palliative care", and "analgesics". We chose these as the main variables among a total of 5,966 variables related to "opioids" and "cancer".

Figure 3

Bibliometric network

# Table 6

Bibliometric network clusters

| Cluster | Keyword | Occurrences |
|---|---|---|
| **1. Opioids** | opioids | 507 |
| | cancer pain | 259 |
| | chronic pain | 173 |
| | pain management | 125 |
| | neuropathic pain | 55 |
| | methadone | 47 |
| | addiction | 37 |
| | chronic non-cancer pain | 26 |
| | naloxone | 25 |
| | breakthrough cancer pain | 23 |
| | tramadol | 18 |
| | ketamine | 17 |
| | pharmacokinetics | 17 |
| | pharmacogenetics | 16 |
| | primary care | 16 |
| | opioid use disorder | 15 |
| | pain assessment | 15 |
| | prescription opioids | 15 |
| | systematic review | 15 |
| **2. Pain** | pain | 259 |
| | cancer | 190 |
| | opioid | 175 |
| | analgesics | 95 |
| | analgesia | 74 |
| | opioid analgesics | 49 |
| | neoplasms | 39 |
| | postoperative pain | 25 |
| | adverse effects | 23 |
| | surgery | 22 |
| | survival | 21 |
| | head and neck cancer | 18 |
| | review | 17 |
| | pain measurement | 16 |
| | mortality | 15 |
| | inflammation | 14 |
| **3. Morphine** | morphine | 101 |
| | fentanyl | 69 |
| | breakthrough pain | 54 |
| | oxycodone | 53 |
| | breast cancer | 32 |
| | treatment | 21 |
| | pharmacoepidemiology | 21 |
| | buprenorphine | 16 |
| | opioid rotation | 15 |
| **4. Palliative Care** | palliative care | 211 |
| | quality of life | 49 |
| | dyspnea | 41 |
| | advanced cancer | 23 |
| | oncology | 22 |
| | end of life | 22 |
| | lung cancer | 19 |
| | hospice | 18 |
| | end-of-life care | 18 |
| | elderly | 18 |
| | symptoms | 16 |
| | palliative | 16 |
| | supportive care | 15 |
| | breathlessness | 14 |

In Figure 5 we show the centrality of the University of Washington and its high collaboration.

In Table 7, we show the centrality of the institutions, that shown in Figure 4, is presented as a bibliometric network. We show the collaboration between institutions, with prominent participation of the University of Washington, Maddalena Cancer Center, Johns Hopkins University, MD Anderson Cancer Center of the University of Texas and the University of Wisconsin-Madison. We noticed a strong collaboration between the Maddalena Cancer Center and the University of Palermo.

Table 7

Centrality of institutions

| Institution | Centrality |
| --- | --- |
| Univ Washington | 346.53 |
| La Maddalena Canc Ctr | 203.68 |
| Johns Hopkins Univ | 195.23 |
| Univ Texas Md Anderson Canc Ctr | 159.12 |
| Univ Wisconsin | 150.51 |
| Norwegian Univ Sci And Technol | 98.80 |
| Stanford Univ | 50.34 |
| Harvard Univ | 49.57 |
| St Olavs Univ Hosp | 47.35 |
| Harvard Med Sch | 35.98 |
| Univ Roma La Sapienza | 27.94 |
| Univ Melbourne | 24.53 |
| Mem Sloan Kettering Canc Ctr | 21.25 |
| Univ Queensland | 20.47 |
| Temple Univ | 17.94 |
| Univ Kentucky | 12.58 |
| Univ Palermo | 10.67 |

Table 8

Collaboration and articles per country

| Country | Articles | Collaboration |
|---|---|---|
| USA | 1636 | 246 |
| Italy | 416 | 58 |
| Germany | 344 | 78 |
| United Kingdom | 325 | 81 |
| Canada | 214 | 46 |
| Australia | 206 | 64 |
| Japan | 191 | 8 |
| China | 116 | 17 |
| France | 105 | 18 |
| Spain | 105 | 15 |

The high scientific production and collaboration of the USA is shown in Table 8, with 1,636 published articles and collaboration with other countries in 246 articles, which is specifically shown in Figure 5 and the University of Washington with 138 articles. The USA is also found with 53,616 citations, followed by Italy with 10,172 citations, and the United Kingdom with 10,098 citations of scientific articles with the keywords "opioid" and "cancer".

The following authors are the most relevant, as related with the search of "opioid" and "cancer": Mercadante S. with 193 artículos, Bruera E. with 133 artículos, Radbruch L. with 69 artículos, Kaasa S. with 64 artículos y Casuccio A. with 62 artículos. The article "Use of opioid analgesics in the treatment of cancer pain: evidence-based recommendations from the EAPC" has 300 citations.

Figure 4

Analysis of variables

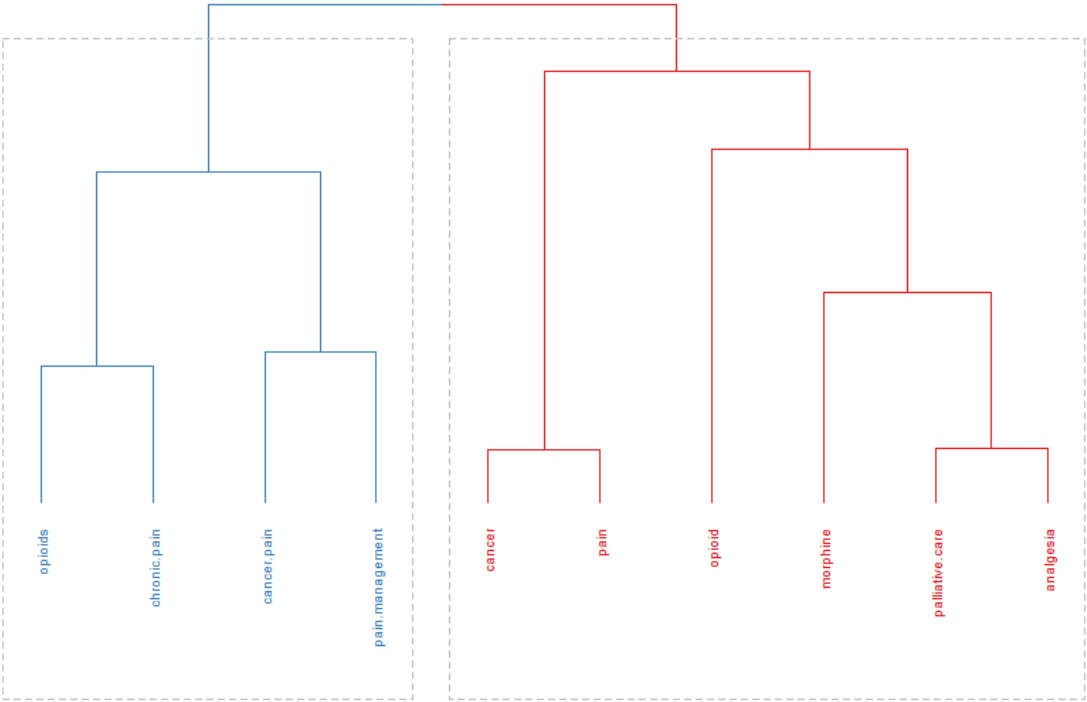

Figure 5

Bibliographic Institution Network

Figure 6

University of Washington Centrality

**Conclusions**

We suggest that the maturity of opioids is increasingly associated with the pain provoked by cancer and with the chronic pain associated with non-cancerous illness, as shown by the results of the scientific articles analyzed with the bibliometric method.

The maturity of the use of opioids was measured by the increase of use in the treatment of pain caused by cancer.

This bibliometric method allowed us to construct an intellectual structure of opioids and its relationship with cancer.  This way we analyzed the maturity process of the use of opioids, beyond the fact that opioids can be addictive, if we consider indistinct variables to measure the maturity of opioids.

We identified the main variables, as they evolve in time and its dynamism, tendency and growth lead us to conclude that opioid use has reached maturity by the different studies that show that they are used in the treatment of pain caused by cancer and non-cancerous illness.

We showed the main indicators of the collaboration and international participation to show that opioids have a strong link with cancer and pain.

**References**

1. Park, S. S. (2010). Cancer Pain Management-Opioids. Journal of the Korean Medical Association, **53**(3), 250–257. https://doi.org/10.5124/jkma.2010.53.3.250

2. Kalso, E., Edwards, J. E., Moore, R. A., & McQuay, H. J. (2004). Opioids in chronic non-cancer pain: systematic review of efficacy and safety. PAIN, **112**(3), 372–380. https://doi.org/10.1016/j.pain.2004.09.019

3. Manchikanti, L., Helm II, S., Fellows, B., Janata, J. W., Pampati, V., Grider, J. S., & Boswell, M. V. (2012). Opioid Epidemic in the United States. PAIN PHYSICIAN, **15**(3, S, SI), ES9–ES38. https://europepmc.org/article/med/22786464

4. Cherny, N., Ripamonti, C., Pereira, J., Davis, C., Fallon, M., McQuay, H., Mercadante, S., Pasternak, G., Ventafridda, V., & Net, E. A. P. C. (2001). Strategies to manage the adverse effects of oral morphine: An evidence-based report. JOURNAL OF CLINICAL ONCOLOGY, **19**(9), 2542–2554. https://doi.org/10.1200/JCO.2001.19.9.2542

5. Portenoy, R. K. (1996). Opioid therapy for chronic nonmalignant pain: A review of the critical issues. Journal of Pain and Symptom Management, **11**(4), 203–217. https://doi.org/10.1016/0885-3924(95)00187-5

6. Zenz, M., Strumpf, M., & Tryba, M. (1992). Long-Term Oral Opioid Therapy in Patients with Chronic Nonmalignant Pain. Journal of Pain and Symptom Management, **7**(2), 69–77. https://doi.org/10.1016/0885-3924(92)90116-Y

7. Katz, N. (2002). The impact of pain management on quality of life. Journal of Pain and Symptom Management, **24**(1, S), S38–S47. https://doi.org/10.1016/S0885-3924(02)00411-6

8. Fajardo D., Durán L., Moreno L., Ochoa H. & Castaño V.M. (2014) Liposomes vs. metallic nanostructures: differences in the process of knowledge translation in cancer, Int. J. Nanomed. **9**, 2627. https://doi.org/10.2147/IJN.S62315

9. Fajardo D., Ortega J. & Castaño V.M. (2015) Hegemonic structure of basic, clinical and patented knowledge on Ebola research: A US Army reductionist initiative, J. Transl. Med. **13**, 124. https://doi.org/10.1186/s12967-015-0496-y

10. Aguado C. & Castaño V.M. (2020) Translational Knowledge Map of COVID-19. Cornell University. arXiv:2003.10434 [cs.DL]. https://arxiv.org/abs/2003.10434